# Innovation, Liquidity and Banks

Gerasimos Lianos[1] and Igor Sloev[2]

## ABSTRACT

We present an infinite-horizon equilibrium model for the allocation of investment between illiquid innovation and liquid storage. There are overlapping generations of individuals subject to an idiosyncratic liquidity shock. Banks transform offer liquidity insurance to households and decide on the allocation of funds into innovation and storage. There is a liquidity constraint on the part of banks, binding under the assumptions on returns. There is a financial accelerator of net worth for the composition of investment. Financial leverage hampers innovation due to a debt overhang effect. Due to optimal liquidity insurance and the overlapping generations nature of the economy, however, a higher probability of early liquidity shock fosters innovation, due to lower interest rates on debt relieving debt overhang. In the case of low return o storage compared to pledgeable return on innovation, the equilibrium is stable. In the case of high storage return compared to innovation pledgeable return, however, the equilibrium is unstable, giving rise to exploding investment in innovation.

[1] Higher Colleges of Technology, Abu Dhabi College. Sheikh Mohammed Bin Rashid Al Maktoum St, Al Mizn, Baniyas North, Abu Dhabi, UAE. glianos@hct.ac.ae
[2] European University at St Petersburg. sia@eu.spb.ru

# Innovation, Liquidity and Banks

## 1. Introduction

We live in a period of stunning technological change, with exorbitant valuations in some technology sectors in the lieu of increased uncertainty and suppressed financial returns. How can long-term investment in innovation be reconciled with increased concerns about short-term risks?

Our purpose is to examine allocation of investment between innovation and conventional investment (storage) in an environment of present liquidity shocks. We do this using a version of the economy of Bernanke and Gertler (1987) and Diamond and Dybvig (1983), under different assumptions about investment returns.

## 2. The Model

The economy consists of households and bankers. Households form overlapping generations of individuals of size *1*. An individual lives for three periods. An individual born in period t receives endowment $w_t$ at date *t* and consumes either in period *t+1* (*type-I* consumer), or in period *t+2* (*τype-II* consumer). An individual has exogenous probability $\alpha$ to be a *type-I* consumer. Newly born individuals deposit $d_t$ at banks before learning their type.

The ex-post date-*t+1* and date-*t+2* utilities of an individual born at $t$ are:

$$u(c_{t+1}^{I}) = log\ c_{t+1}^{I};\ u(c_{t+2}^{II}) = log\ c_{t+2}^{III}.$$

The ex-post budget constraint of a *type-I* individual born at *t*:

$$c_{t+1}^{I} = R^1\, d_t \tag{1}$$

where $R^1$ is the gross return on deposit for *type-I* of consumers.

The ex-post constraint a *type-II* individual born in period t:

$$c_{t+1}^{II} = R^2\, d_t \tag{2}$$

where $R^2$ is the gross return on deposits for *type-II* consumers.

Household endowments are given to bank as deposits.

$$d_t = w_t \tag{3}$$

Banks are infinite-lived, risk-neutral and of measure of *1*. Banks have exogenous equity $e_t$ in period *t*. There are two technologies: (a) a long-term technology (innovation) whose returns are only partly pledgeable and (b) a short-term technology (storage). Both technologies offer safe returns in the sense that there is no randomness in the returns. A bank invests $i_t$ in period *t* in innovation that generates gross return of $Ai_t$ in period *t+2*. Investment of $s_t$ in storage in period *t* yields gross return of $Rs_t^b$ in period *t+1*. The flow-of-funds constraint is:

$$d_t + e_t = i_t + s_t \tag{4}$$

Bankers can divert fraction $\theta$ of the returns to the innovation. Hence, only $(1-\theta)A$ is pledgeable to the depositors. There is no diversion in storage. We make the following assumption about technology returns.

We consider two alternative assumptions about the returns.

### Assumption 1

$$(1-\theta\ )A < R < \beta A \quad \text{(A1)}$$

### Assumption 2

$$\beta(1-\theta\ )A < R < (1-\theta)A < \beta A \quad \text{(A2)}$$

Both (A1) and A(2) imply that neither innovation nor storage dominate the other. However, (A1) corresponds to an economy with high return on storage or equivalently with low pledgeable return on innovation. (A2) corresponds to the opposite. We will consider the dynamic behavior of the economy under the two alternative assumptions.

The bank's intertemporal profit is the difference of the returns from innovation and storage and the withdrawals on deposits.

$$\pi_t \ = \ Ai_{t-2} + Rs_{t-1} \ - \alpha R^1 d_{t-1} - (1-a)R^2 d_{t-2} \quad (5)$$

A liquidity constraint must hold at all periods. The liquidity constraint means that the bank has always enough resources to meet withdrawals, which prevents bank runs.

$$(1-\theta)Ai_{t-1} + Rs_t \geq \alpha\ R^1 d_t + (1-\alpha)R^2 d_{t-1} \quad (6)$$

### The Equilibrium

The bank chooses sequences of innovation, storage and deposits $\{i_t, s_t, d_t\}_{t=0}^{\infty}$ to maximize the discounted sum of profits.

$$\sum_{t=0}^{\infty} \beta^t \pi_t$$

subject to (1) and (3), given the sequences of prices.

Given that there are no restrictions on storage, deposits are exogenous to the bank. The Lagrangian function of the maximization problem is:

$$\Lambda_0 = \sum_{t=0}^{\infty} \beta^t \pi_t + \beta^{t+1}\lambda_t[(1-\theta)Ai_{t-1} + Rs_t - \alpha R^1 d_t - (1-\alpha)R^2 d_{t-1}]$$

where $\lambda_t$ is the Lagrange multiplier at time $t$, representing the shadow value of the liquidity constaint of period *t+1*.

At an interior equilibrium of the bank's optimization problem with respect to $i_t$ and $s_t$ are:

$$\beta[1 + \lambda_{t+1}(1-\theta)]A = (1+\lambda_t)\alpha R^1 + \beta(1+\lambda_{t+1})(1-\alpha)R^2 \quad (7)$$

$$(1+\lambda_t)R = (1+\lambda_t)\alpha R^1 + \beta(1+\lambda_{t+1})(1-\alpha)R^2 \quad (8)$$

Eq. (5) is the optimality condition with respect to innovation. It says that the total expected marginal benefit of investing in a loan (left side) be equal to the total expected marginal cost of the investment in terms of period *t+1* consumption (right side). The total marginal benefit has two components: the direct expected return on the projects and the indirect expected benefit of having more liquidity in period *t+2*. The expected marginal benefit is discounted to the period-*(t+1)* value. The total expected marginal cost of investing in the long-term project equals the total expected cost of raising one additional unit of deposits. This is comprised by the direct and indirect cost of raising one unit of deposits. The direct cost consists of the expected return on one unit of demand deposits raised in period t and withdrawn in period *t+1*, plus the expected return on one unit of demand deposits raised in period *t* and withdrawn in period *t+2*, discounted to period *t+1*. The indirect cost equals the expected return on one unit of demand deposits raised in period t and withdrawn in period *t+1*, for a bank constrained in period *t+1*, and the expected return on one unit of demand deposits raised in period t and withdrawn in period *t+2*, discounted to period *t+1*, for a bank constrained in period *t+2*.

Eq. (6) is the optimality condition with respect to storage. It requires that the total marginal expected benefit of investing one unit of endowments in storage equals the total cost (right side) of financing it through consumer deposits. The total expected marginal benefit is comprised by the direct benefit, the rate of return to storage, plus the indirect benefit of having one additional unit of liquidity in period t+1 for a constrained intermediary. The total expected marginal cost of one unit of storage is the cost of raising one unit of deposits.

The bank provides optimal liquidity insurance by offering the optimal menu of deposits to early and late consumers. The equilibrium returns on deposits maximize individual's expected utility subject to the ex-post budget constraints and the non-arbitrage condition.

**Lemma 1.** The equilibrium returns on deposits $(R^1, R^2)$ are given by:

$$R^1 = \frac{1}{\alpha+\rho(1-\alpha)} R \qquad (9)$$

$$R^2 = \left(\frac{\rho}{\beta}\right)\left[\frac{1}{\alpha+\rho(1-\alpha)}\right]\left(\frac{1+\lambda_t}{1+\lambda_{t+1}}\right) R \qquad (10)$$

Substituting for $R^1$, $R^2$ and $s_t$ from (1) into (3), we get the law of motion for innovation:

$$i_t = \frac{1}{R}\left[(1-\theta)Ai_{t-1} + Re_t + \frac{\rho(1-\alpha)}{\alpha+\rho(1-\alpha)}Rw_t - \frac{\rho(1-\alpha)}{\beta[\alpha+\rho(1-\alpha)]}\left(\frac{1+\lambda_t}{1+\lambda_{t+1}}\right)Rw_{t-1}\right] \tag{11}$$

From (5) and (6), we get:

$$\lambda_{t+1} = \frac{R}{\beta(1-\theta)A}\lambda_t + \frac{R-\beta A}{\beta(1-\theta)A} \tag{12}$$

## The Steady-State Equilibrium

Lemma 2. The liquidity constraint is binding at the equilibrium.

$$\lambda^* > 0$$

The steady-state Lagrange multiplier is:

$$\lambda^* = \frac{R-\beta A}{\beta(1-\theta)A-R} \tag{13}$$

$$i^* = \frac{R}{R-(1-\theta)A}\left[e - \frac{\rho(1-\alpha)}{\alpha+\rho(1-\alpha)}\left(\frac{1-\beta}{\beta}\right)w\right] \tag{14}$$

$$s^* = e + w - i^* \tag{15}$$

Let $m(l_t) = \frac{1-\delta l_t}{1+l_t}$; where $l = w/e$ is the bank's leverage ratio. The innovation-to-investment ratio at the steady state:

$$\left(\frac{i}{s}\right)^* = \frac{m(l)}{1-q-m(l)} \tag{16}$$

We have the following propositions regarding the stability of the steady state, corresponding to assumptions (A1) and (A2). [Proofs in the Appendix]

Proposition 1.

Under assumption (A1), the system of equations (11) and (12) give rise to a saddle stable path. Along the saddle path each date, the Lagrange multiplier must be at the steady-state value $\lambda^*$.

Proposition 2.

Under the assumption (A2), the system of equations (11) and (12) give rise to an explosive source.

Let: $q = \frac{(1-\theta)A}{R}$; $\delta = \frac{\rho(1-\alpha)}{\alpha+\rho(1-\alpha)}\left(\frac{1-\beta}{\beta}\right)$. Along the saddle path, innovation is:

$$i_t = qi_{t-1} + e - \delta w \tag{17}$$

Letting $x_t = \left(\frac{i_t}{s_t}\right)$, the innovation-to-investment ratio is:

$$x_t = \frac{m(l)+[m(l)+q]x_{t-1}}{1-m(l)+[1-m(l)-q]x_{t-1}} \tag{18}$$

From (16) and (18), we obtain:

Proposition 3.

The innovation-to-investment ratio decreases with the bank's leverage in the steady state and transitory equilibrium.

$\frac{\partial(x^*)}{\partial l} < 0$; $\frac{\partial(x_t)}{\partial l} < 0$

## 3. Discussion

We have used an instrumental definition of innovation: long-term investment, burdened by agency costs (assuming the form of Hart and Moore (1994) and Kiyotaki and Moore (1997)). The approach to innovation versus storage as long-term versus short-term investment is in the spirit of Aghion *et. al.* (2010). Moreover, innovation has the feature of corresponding to risky, intangible, and non-collaterizable assets (Hall & Lerner (2010); Brown *et. al.* (2013)). The agency cost approach followed here is equivalent to standard moral hazard formulations (e.g. Tirole (2006)). Our approach is also mathematically equivalent to a conceptualization of innovation as long-term risky investment vis-à-vis a safe investment (storage). In this formulation (Bernanke and Gertler (1987)), we would have:

$$A = E(R^I); (1-\theta)A = R^{min}$$

with $E(R^I)$ the expected return to innovation and $R^{min}$ the minimum return to innovation.

Our model contributes to the innovation literature by giving a tractable dynamic framework in which capital structure affects the allocation between innovation and storage (conventional investment). The key mechanism is the presence of a debt-induced innovation wedge, which makes leverage reduce the innovation-to-storage ratio for constant total financing. The innovation-to-investment ratio $i_t/s_t$ is governed by the excess pledgeable return to innovation:

$$q = (1-\theta)A/R$$

and the debt-overhang on innovation:

$$\delta = \{\rho(1-\alpha)/[\alpha + \rho(1-\alpha)]\}\{1-\beta)/\beta\}.$$

In the theory of financial accelerator (Bernanke & Gertler, M. (1989); Bernanke, *et. al*. (1999)), balance sheets matter because net worth affects agency costs, external finance premia and real investment. Our model shows the operation of a similar financial accelerator affecting the composition of investment: leverage decreases the amount of funds directed to innovation. This can be seen in the innovation equation:

$$i_t = qi_{t-1} + e - \delta w$$

in which enters positively, while debt negatively. Debt carries an innovation wedge (Myers (1977); Hart & Moor (1995); Diamond & He (2014)).

An analytical contribution is the tractability of the model. The model expresses the innovation-to-storage ratio as:

$$x_t = \frac{m + (m+q)x_{t-1}}{1 - m + (1 - m - q)x_{t-1}}$$

Our model accords  the empirical literature (e.g. Mann (2018)) that increased pledgeablity of returns (greater fraction of non-divertible assets $1 - \theta$) increases innovation, as it can be seen by the partial derivative:

$$\frac{\partial x^*}{\partial \theta} < 0$$

A property of our model is that leverage can be beneficial for innovation when agency costs are higher (greater fraction of divertible assets $\theta$). This can be seen by the cross-partial derivative of the innovation-to-storage ratio:

$$\frac{\partial^2 x^*}{\partial \theta \partial l} > 0$$

which provides a rational for the empirical fact that debt promotes innovation in young, more risk-ridden firms and industry entrants (Geelen *et. al*. (2022)).

An interesting feature of our model is that greater liquidity preference, in the sense of a greater fraction of households being early-type consumers, (perhaps counterintuitively) enhances innovation. This is due to the dynamic nature of the environment. Regardless of the proportion of early consumers $\alpha$, the size of liquidity demand (withdrawals) is the same:

$$\alpha\, R^1 d + (1 - \alpha) R^2 d$$

A greater fraction of early consumers, however, has the effect of decreasing both interest rates $R^1, R^2$, as

$$R^1 = \left[\frac{1}{\alpha+\rho(1-\alpha)}\right] R;\ R^2 = \left(\frac{\rho}{\beta}\right)\left[\frac{1}{\alpha+\rho(1-\alpha)}\right] R$$

hence reducing the debt-overhang on innovation $\delta$. An implication of this fact is that greater risk of liquidity shock in the presence of optimal liquidity insurance enhances innovation. This can be seen by interpreting $\alpha$ as probability of occurrence of a liquidity shock and $1-\alpha$ as the probability of non-occurrence.

We have considered two profiles for return to storage and pledgeable return to innovation. When pledgeable return to innovation is low, in the sense that: $(1-\theta)A < R$, the equilibrium is stable under temporary shocks. When pledgeable return is high, meaning: $R < (1-\theta)A$, however, even if innovation does not dominate storage (i.e. $\beta(1-\theta)A < R$), a perturbation of the equilibrium gives rise to explosive investment in innovation. Although the case of saddle path stability is usually desirable in macroeconomic models, the case of instability may be useful to think about phenomena such as rational exuberance and asset pricing bubbles. Our model says that shortage of media of storage (high price/low return of media of storage) (Caballero *et. al.* (2017), Holmstrom and Tirole (1998)) or increased ability to securitize illiquid assets may give rise to observed phenomena of explosive dynamics in the investment (and price) of the illiquid assets.

## 4. Conclusion

We present a tractable infinite-horizon equilibrium model for the allocation of investment between illiquid innovation and liquid storage. There are overlapping generations of individuals subject to an idiosyncratic liquidity shock. Banks transform offer liquidity insurance to households and decide on the allocation of funds into innovation and storage. There is a liquidity constraint on the part of banks, binding under the assumptions on returns. There is a financial accelerator of net worth for the composition of investment. Financial leverage hampers innovation due to a debt overhang effect. Due to optimal liquidity insurance and the overlapping generations nature of the economy, however, a higher probability of early liquidity shock fosters innovation, due to lower interest rates on debt relieving debt overhang. In the case of low return on storage compared to pledgeable return on innovation, the equilibrium is stable. In the case of high storage return compared to innovation pledgeable return, however, the equilibrium is unstable, giving rise to exploding investment in innovation.

## APPENDIX

*The Equilibrium Returns to Deposits*

The individual's expected-utility maximization problem at date t:

$$\max_{R^1,R^2} E\{U(c^I_{t+1}, c^{II}_{t+2})\} = \alpha log\ c^I_{t+1} + \rho(1-\alpha) log\ c^{II}_{t+2}\}$$

subject to the constraints:

$$c^I_{t+1} = R^1 d_t;\ c^{II}_{t+1} = R^2 d_t;\ d_t = w_t;\ (1+\lambda_t)R = (1+\lambda_t)\alpha R^1 + \beta(1+\lambda_{t+1})(1-\alpha)R^2$$

PROOF OF LEMMA 1. The Lagrangian formulation of the household´s optimization problem (using (1), (2) and (3)) is given by:

$$M = \max_{R^1,R^2} log[(R^1 w_t)^\alpha (R^2 w_t)^{1-\alpha}] + \mu[(1+\lambda_t)R - (1+\lambda_t)\alpha R^1 - \beta(1+\lambda_{t+1})(1-\alpha)R^2]$$

The first-order conditions of the problem give:

$$R^2 = \frac{\rho}{\mu\beta(1+\lambda_{t+1})} \quad (19)$$

$$R^1 = \frac{1}{\mu(1+\lambda_t)} \quad (20)$$

Combining (19) and (20), we get:

$$R^1 = \frac{\beta(1+\lambda_{t+1})}{\rho(1+\lambda_t)} R^2 \tag{21}$$

Substituting for $R^1$ from (21) into the non-arbitrage condition (7), we obtain (9). Substituting for $R^2$ from (9) into (21) we get (8).

*The Steady-State Equilibrium*

The steady-state value for $i^*$ satisfies: $[R-(1-\theta)A]i^* = R\left\{e - \left[\frac{\rho(1-\alpha)}{\alpha+\rho(1-\alpha)}\right]\left(\frac{1-\beta}{\beta}\right)w\right\}$

For $i^*$ to be positive a necessary and sufficient condition is: $\frac{e}{w} > \left(\frac{1-\beta}{\beta}\right)\left[\frac{\rho(1-\alpha)}{\alpha+\rho(1-\alpha)}\right]$

The dynamics of the economy are given by the system of (10), (11), and (12), with the initial conditions for investment $i_0$ and the Lagrange multiplier $\lambda_0$. Since: $\frac{R}{\beta(1-\theta)A} > 1$ and $R-\beta A < 0$, for each profile of expected returns $(A, R)$ there is a unique $\lambda^*$, for which for $\lambda_t < \lambda^*$ the Lagrange multipliers tend to zero and for $\lambda_t > \lambda^*$ the Lagrange multipliers tend to +∞.

The steady state $(\lambda^*, i^*, s^*)$ is given by the system:

$$\lambda^* = \frac{R-\beta A}{\beta(1-\theta)A-R},\ i^* = \frac{R}{R-(1-\theta)A}\left[e - \frac{\rho(1-\alpha)}{\alpha+\rho(1-\alpha)}\left(\frac{1-\beta}{\beta}\right)w\right],\ s^* = e+w-i^*$$

PROOF OF LEMMA 2. Under (A1) and (A2), $\lambda^* = \frac{R-\beta A}{\beta(1-\theta)A-R} > 0$

*The Stability of the Steady-State Equilibrium*

Let $\hat{\lambda}_t = \lambda_t - \lambda^*$, $\hat{\imath}_t = i_t - i^*$, and $\hat{s}_t = s_t - s^*$ be the deviations from the steady state. It is:

$$\hat{\lambda}_{t+1} = \frac{R}{\beta(1-\theta)A}\hat{\lambda}_t.$$

Linearizing $\left(\frac{1+\lambda_t}{1+\lambda_{t+1}}\right)$, we have: $\left(\frac{1+\lambda_t}{1+\lambda_{t+1}}\right) = 1 + \frac{1}{1+\lambda^*}\left(\hat{\lambda}_t - \hat{\lambda}_{t+1}\right) = 1 + \frac{1-b}{1+\lambda^*}\hat{\lambda}_t$, $b = \frac{R}{\beta(1-\theta)A}$,

which gives: $\hat{\imath}_t = \frac{(1-\theta)A}{R}\hat{\imath}_{t-1} + \frac{1}{\beta}\left(\frac{\rho(1-\alpha)}{\alpha+\rho(1-\alpha)}\right)\left(\frac{b-1}{1+\lambda^*}\right)w\hat{\lambda}_t$

Let: $a \equiv \frac{(1-\theta)A}{R}$ and $\kappa \equiv \frac{1}{\beta}\left(\frac{\rho(1-\alpha)}{\alpha+\rho(1-\alpha)}\right)\frac{b-1}{1+\lambda^*}w$. Then the linearized dynamic system is:

$\begin{pmatrix} \hat{\imath}_t \\ \hat{\lambda}_{t+1} \end{pmatrix} = J\begin{pmatrix} \hat{\imath}_{t-1} \\ \hat{\lambda}_t \end{pmatrix}$ where J is the Jacobian matrix: $J = \begin{bmatrix} a & \kappa \\ 0 & b \end{bmatrix}$, with the eigenvalues:

$\mu_1 = a = \frac{(1-\theta)A}{R}$; $\mu_2 = b = \frac{R}{\beta(1-\theta)A}$.

*Case 1.* Under the assumption (A1), $\mu_1 < 1; \mu_2 = \frac{R}{\beta(1-\theta)A} > \frac{(1-\theta)A}{\beta(1-\theta)A} = \frac{1}{\beta} > 1$

Therefore the steady state has one stable eigenvalue and one unstable eigenvalue. Hence, the steady state is a saddle point.

To find the eigenspace associated with the stable eigenvalue $\mu_1$, we solve: $(J - \mu_1 I)v = 0$

$\begin{pmatrix} \mu_1 & \kappa \\ 0 & \mu_2 \end{pmatrix} - \begin{pmatrix} \mu_1 & 0 \\ 0 & \mu_1 \end{pmatrix} = \begin{pmatrix} 0 & \kappa \\ 0 & \mu_2 - \mu_1 \end{pmatrix}$. Letting: $v = \begin{pmatrix} v_i \\ v_\lambda \end{pmatrix}$, the eigenvector condition is

$\begin{pmatrix} 0 & \kappa \\ 0 & \mu_2 - \mu_1 \end{pmatrix}\begin{pmatrix} v_i \\ v_\lambda \end{pmatrix} = \begin{pmatrix} 0 \\ 0 \end{pmatrix}$

The second row gives: $(\mu_2 - \mu_1)v_\lambda = 0$. Since $\mu_2 > 1 > \mu_1$, we have: $\mu_2 - \mu_1 \neq 0$. Therefore: $v_\lambda = 0$. The first component $v_i$ is unrestricted. Hence, every stable eigenvector has the form:

$v = \begin{pmatrix} v_i \\ 0 \end{pmatrix}$ and the stable eigenspace is: $E^s = \left\{\begin{pmatrix} \hat{\imath}_{t-1} \\ \hat{\lambda}_t \end{pmatrix} : \hat{\lambda}_t = 0\right\}$

Equivalently, $\hat{\lambda}_t = 0$ is the condition for being on the stable eigenspace. Thus, for convergence to the steady state, the multiplier must jump immediately to its steady-state value: $\lambda_t = \lambda^*$. Then innovation follows: $\hat{\imath}_t = \frac{(1-\theta)A}{R}\hat{\imath}_{t-1}$ and since $0 < \frac{(1-\theta)A}{R} < 1$, we have: $\hat{\imath}_t \to 0$. As $s_t = e + w - i_t$, the investment deviation satisfies: $\hat{s}_t = -\hat{\imath}_t$. Therefore, along the stable saddle path, $\hat{s}_t \to 0$.

*Case 2.* Under Assumption (A2), $\mu_1 > 1$; $\mu_2 > 1$. Hence, the equilibrium is an unstable source.

## References

Aghion, P., Angeletos, G-M, Banerjee, A. and K. Manova (2010). "Volatility and growth: Credit constraints and the composition of investment", Journal of Monetary Economics, Elsevier, vol. 57(3), 246-265.

Bernanke B. and M. Gertler (1987), “Banking in Macroeconomic Equilibrium”, in New Approaches to Monetary Economics, edited by W. A. Barnett and K. J. Singleton, Massachusetts, Cambridge Press.

Bernanke, B., & Gertler, M. (1989). “Agency Costs, Net Worth, and Business Fluctuations.” American Economic Review, 79(1), 14–31.
Bernanke, B. S., Gertler, M., & Gilchrist, S. (1999). “The Financial Accelerator in a Quantitative Business Cycle Framework.” In J. B. Taylor & M. Woodford, eds., Handbook of Macroeconomics, Vol. 1C, 1341–1393. Amsterdam: Elsevier. DOI: 10.1016/S1574-0048(99)10034-X

Brown, J. R., Martinsson, G., & Petersen, B. C. (2013). “Law, Stock Markets, and Innovation.” Journal of Finance, 68(4), 1517–1549. DOI: 10.1111/jofi.12040.

Caballero, R., Farhi E., and P.O. Gourinchas. 2017. "The Safe Assets Shortage Conundrum." *Journal of Economic Perspectives* 31 (3): 29–46.DOI: 10.1257/jep.31.3.29

Diamond D. and P. Dybvig (1983), “Banks Runs, Deposit Insurance and Liquidity”, Journal of Political Economy, vol. 91, no. 3, pp. 401-419.

Diamond, D. W., & He, Z. (2014). “A Theory of Debt Maturity: The Long and Short of Debt Overhang.” Journal of Finance, 69(2), 719–762. DOI: 10.1111/jofi.12118.

Geelen, T., Hajda, J., & Morellec, E. (2022). “Can Corporate Debt Foster Innovation and Growth?” Review of Financial Studies, 35(9), 4152–4200. DOI: 10.1093/rfs/hhab129.

Hart O. & J. Moore, (1994) “A Theory of Debt Based on the Inalienability of Human Capital”, The Quarterly Journal of Economics, 109(4), pp. 841-879

Hart, O., & Moore, J. (1995). “Debt and Seniority: An Analysis of the Role of Hard Claims in Constraining Management.” American Economic Review, 85(3), 567–585.

Holmström, B., & Tirole, J. (1998). Private and Public Supply of Liquidity. Journal of Political Economy, 106(1), 1–40. https://doi.org/10.1086/250001.

Kiyotaki N. & J. Moore (1997). “Credit Cycles”. Journal of Political Economy, 105(2), pages 211-248.

Hall, B. H., & Lerner, J. (2010). “The Financing of R&D and Innovation.” In B. H. Hall & N. Rosenberg, eds., Handbook of the Economics of Innovation, Vol. 1, 609–639. Amsterdam: Elsevier. DOI: 10.1016/S0169-7218(10)01014-2.

Mann, W. (2018). “Creditor Rights and Innovation: Evidence from Patent Collateral.” Journal of Financial Economics, 130(1), 25–47. DOI: 10.1016/j.jfineco.2018.07.001

Myers, S. C. (1977). “Determinants of Corporate Borrowing.” Journal of Financial Economics, 5(2), 147–175. DOI: 10.1016/0304-405(77)90015-0.

Tirole, J. (2006). “The Theory of Corporate Finance”. Princeton, N.J.: Princeton University Press.